\documentclass[12pt,preprint]{aastex}

\shorttitle{The Unique Na:O Abundance Distribution in NGC 6791}
\shortauthors{Geisler et al.}

\begin{document}

\title{The Unique Na:O Abundance Distribution in NGC 6791:\\
The First Open(?) Cluster with Multiple Populations}

\author{D. Geisler, S. Villanova}
\affil{Departamento de Astronomia, Casilla 160-C, 
       Universidad de Concepcion, Chile}
\email{dgeisler,svillanova@astro-udec.cl}

\author{G. Carraro}
\affil{European Southern Observatory, Alonso de Cordova 3107, Casilla 19001, Santiago 19, Chile}
\email{gcarraro@eso.org}

\author{C. Pilachowski}
\affil{Dept of Astronomy, Indiana University,  727 East Third Street, Bloomington,
IN 47405-7105, USA}
\email{catyp@astro.indiana.edu}

\author{J. Cummings}
\affil{Departamento de Astronomia, Casilla 160-C, Universidad de Concepcion, Chile}
\email{jcummings@astro-udec.cl}

\author{C. I. Johnson} 
\affil{Dept. of Physics and Astronomy, UCLA, 430 Portola Plaza,  Los Angeles, CA 90095-1547, USA}
\email{cijohnson@astro.ucla.edu}
 \altaffilmark{1} \altaffiltext{1}{National Science Foundation Astronomy and
Astrophysics Postdoctoral Fellow}
\altaffilmark{2} \altaffiltext{2}{Visiting Astronomer, Kitt Peak National
Observatory, National Optical Astronomy Observatories, which is operated by
the Association of Universities for Research in Astronomy, Inc. (AURA) under
cooperative agreement with the National Science Foundation.  The WIYN
Observatory is a joint facility of the University of Wisconsin--Madison,
Indiana University, Yale University, and the National Optical Astronomy
Observatory.}


\author{F. Bresolin}
\affil{Institute for Astronomy, University of Hawaii, 2680 Woodlawn Drive, Honolulu, HI 96822, USA}
\email{bresolin@IfA.Hawaii.Edu}

\begin{abstract}
Almost all globular clusters investigated exhibit a spread in their light 
element abundances, the most studied being a Na:O anticorrelation.
In contrast, open clusters show a homogeneous
composition and are still regarded as Simple Stellar Populations. 
The most probable reason for this difference is that globulars had 
an initial mass high enough to retain primordial gas and ejecta 
from the first stellar generation and thus formed a second generation 
with a distinct composition, an initial mass exceeding that of open clusters.
NGC 6791 is a massive open cluster, and warrants a detailed search for chemical inhomogeneities.
We collected high resolution, high S/N spectra of 21 members covering a wide range of evolutionary
status and measured their Na, O and Fe content.   
We found  [Fe/H]=+0.42$\pm 0.01$, in good agreement with previous values, and no evidence 
for a spread. However,
the Na:O distribution is completely unprecedented. It becomes the first open cluster to
show intrinsic abundance variations that cannot be explained by mixing, and thus the first
discovered to host multiple populations. It is also the first star cluster to
exhibit two subpopulations in the Na:O diagram with one being
chemically homogeneous  while the second  has an
intrinsic spread that follows the anticorrelation  so far displayed only by globular clusters.

NGC~6791 is unique in many aspects, displaying certain characteristics typical of open 
clusters, others more reminiscent of globulars, and yet others, in particular its Na:O behavior investigated here, that are totally unprecedented.
It clearly had a complex and fascinating history. 
\end{abstract}


\section{Introduction}

NGC~6791 is a truly unique object in the Galaxy. Since the first in-depth analysis by
\cite{Ki65}, this cluster was recognized as being both very massive (for an
open cluster (OC)) as well as very old.
Despite being perhaps the oldest OC, with an age of $\sim$8 Gyr \citep{Ca06}, its metallicity
is among the highest of any cluster known. 
Indeed, the initial investigation of its
metallicity yielded +0.75, far exceeding that of any other object
and earning it the label of Super Metal Rich \citep{ST71}.
Subsequent investigations have settled on a somewhat lower but still extreme value of $\sim+0.4$.
The combination of large age and abundance places the cluster
in a unique location in the
age-metallicity relation for the disk.
\citet{Ca06} even suggested a possibly extraGalactic origin for NGC 6791, which would make it
even more exceptional. 
Recently, \citet{Tw11} suggested it might have an age
spread of a Gyr, another extraordinary quality if correct.
It is one of only a very few OCs to show the infamous second parameter problem,
with both a red clump as well as stars on the red and extended blue ends of
a horizontal branch \citep{Pl11,Bu12}. 

The entire field of globular cluster (GC) research has recently undergone a  paradigm shift,
driven by the discovery that they are surprisingly
complex objects, formed by  multiple instead of Simple Stellar Populations, as previously believed. 
All of them so far studied in  detail show at least a  spread in their light element
content, 
the most evident being the
spread in Na and O,  which are anti-correlated \citep{Ca09,Ca10,Gr12}.  \citet{Ca10}
have even argued for a new, chemical definition of a GC as any
object which displays a Na:O anticorrelation. 

\begin{figure}[!ht]
\centering
\includegraphics[angle=0,width=8cm]{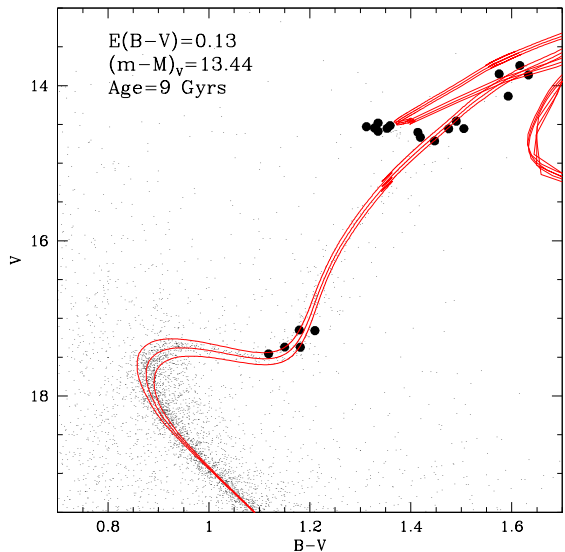}
\caption{CMD of NGC~6791 from \citet{St03} with the observed RGB/RC stars indicated as filled
circles, together with isochrones of 8,9 and 10 Gyrs from \citet{Pi04}.
}
\label{f1}
\end{figure}

The most natural explanation for chemical inhomogeneities is the self-pollution
scenario, where a cluster experiences an extended star formation period, with the younger
population born from an interstellar medium polluted by ejecta from stars of
the older generation that have experienced hot H-burning via p-capture. 
The older generation's composition closely mimics that of similar metallicity halo field stars, while the younger generation is
enhanced in He, N, Na, and Al and depleted in C, O, Ne and Mg.
The material required to form the second generation is retained due to the 
strong gravitational field \citep{De08}. However, there must be a minimum
initial mass required to retain this material. Theoretical limits of the order of $10^5M_\odot$ \citep{Ve10}  and observational values of $\sim4\times10^4M_\odot$ \citep{Ca09}, within the extent of a typical cluster (a few pc), have been estimated for Galactic GCs, while \citet{Mu09} find a limit more like $2\times10^5M_\odot$ for LMC clusters. Unfortunately, these mass estimates are often very uncertain
and, more importantly, refer to the present-day mass. It is
well-known theoretically that a cluster can lose much, most or even all of its
initial mass during its subsequent evolution due to both
internal and external factors \citep{La10}.

On the other hand, OCs so far do not show any spread in chemical
abundances that cannot be attributed to simple, in situ mixing processes \citep{De09}.
This can be explained in the self-pollution scenario
because they formed with an initial mass lower than any GC and below the
above minimum, so they could not retain primordial gas or  ejecta to form a
second generation. 

Given its uniqueness, NGC 6791 has been the subject of many observational studies.
A number of high resolution spectroscopic investigations
have firmly established the metallicity and many details of its chemical
composition \citep{Gr06, Ca06, Or06, Ca07}.
However, no data on both Na and O for more than a few stars are
published. Given the importance of Na and O for examining the formation and 
chemical evolution of clusters, a 
study of these elements in a large sample of stars will fill this gap and constrain 
the nature of this intriguing cluster, which is the aim of this Letter.

\begin{figure*}
\centering
\includegraphics[angle=0,width=15cm]{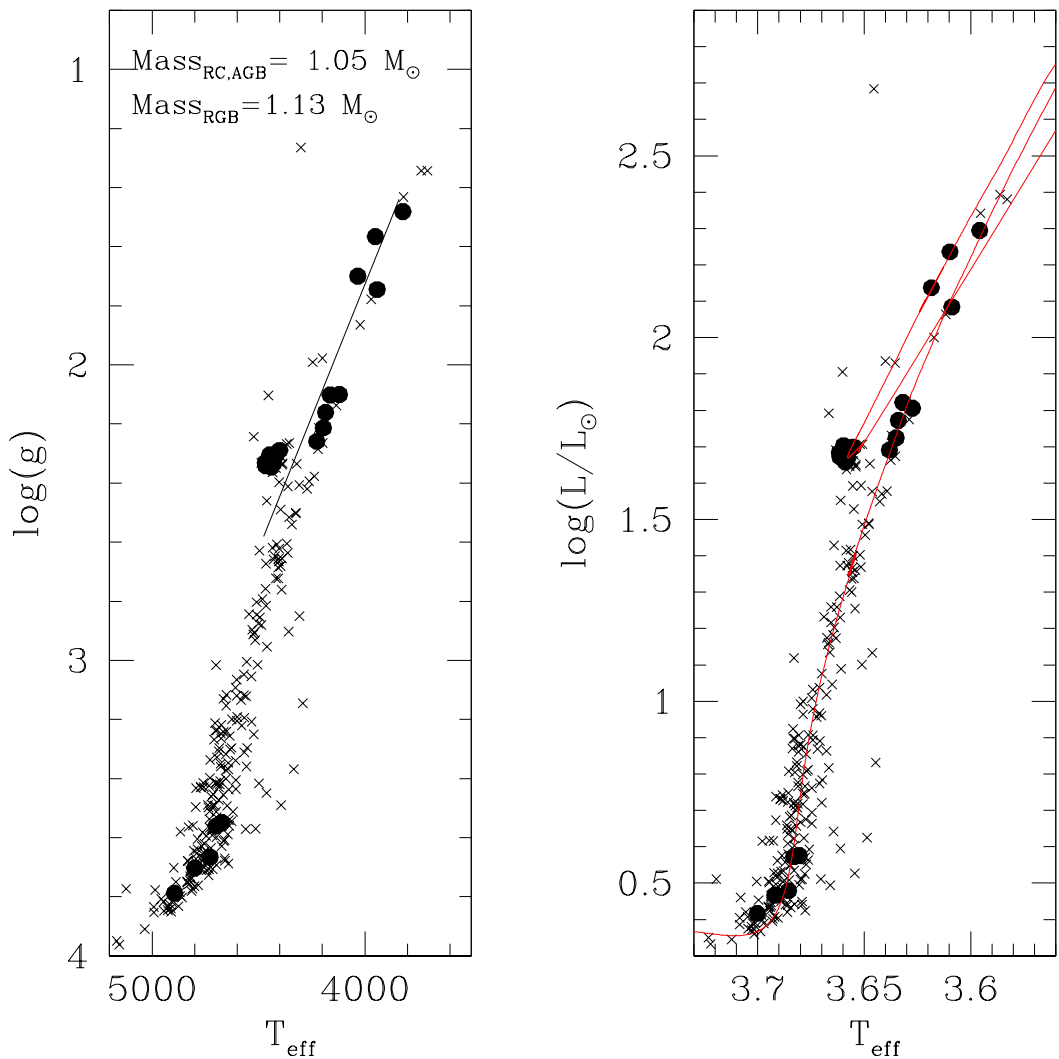}
\caption{Left: log(g) vs T$_{\rm eff}$ for evolved stars in NGC~6791
(crosses) and for our targets (circles). The adopted 
line separates RGB from RC/AGB stars. Masses adopted for RGB and RC/AGB stars
are indicated. Right: log(L/L$_{\odot}$) vs T$_{\rm eff}$ for the
same stars. The red curve is the BASTI isochrone for 9 Gyrs.}
\label{f2}
\end{figure*}

\section{Observations and data reduction}

We employ two independent observations to compile a comprehensive dataset and 
to study as wide a range of evolutionary stages as possible.
One dataset consists of spectra collected 
with HIRES  at the Keck I telescope. 
We observed five stars  
located in the lower part 
of the red giant branch (RGB) (see Fig.~\ref{f1}). Spectra cover the range 3500 - 10000 \AA ~
with a resolution of 45,000. These data are discussed in more detail in \citet{Vi12}.
Our second dataset consists of lower resolution (R$\sim$15,000) spectra obtained 
with the multifiber
Hydra spectrograph at the WIYN telescope.
We observed 19 stars 
located in the upper part 
of the RGB, the red clump (RC) and the asymptotic giant branch (AGB) 
(Fig.~\ref{f1}). 
Spectra cover the range 6050-6375 \AA\ .
Targets were selected on the basis of extensive  B,V,I photometry \citep{St03}. We cross-correlated Stetson's catalogue with 2MASS to obtain J,H,K$_s$ magnitudes.  
Table~\ref{t2} gives optical photometry of the members.

Data were reduced using IRAF\footnote  {IRAF is distributed by the National
Optical Astronomy Observatory, which is operated by the Association of
Universities for Research in Astronomy, Inc., under cooperative agreement
with National Science Foundation}, including bias subtraction, flat-field
correction, wavelength calibration, sky subtraction, spectral rectification,
and combination. Cosmic rays were removed using the program from \citet{VD01}.
HIRES spectra have a typical S/N of
$\sim$50 at 6300 \AA, while Hydra spectra have S/N of
$\sim$70 at this wavelength.

Radial velocities were measured by the {\it fxcor} package in IRAF,
using a synthetic spectrum as template. 
The mean heliocentric value is -44.6$\pm$0.5 km/s, while
the dispersion is 2.3$\pm$0.4 km/s,  in good agreement with published values, e.g. \citet{Ge88,Ca06,
GC12}.
Three of our sample have very different velocities from the 
mean and were rejected as non-members. 
On the basis of radial velocity and metallicity, 
we conclude that all of the other targets are definite cluster members.

\section{Abundance analysis}

The Fe abundances were obtained from the equivalent widths (EWs). 
The main problem was the continuum determination, due to
the very high metallicity. We solved this by comparing our 
spectra with a synthetic one having the
mean atmospheric parameters of the targets and  using as
continuum only those portions of the observed spectra where the corresponding
synthetic spectrum was $\leq1\%$ below the theoretical continuum.
For O and Na, whose lines are affected by blending, including by molecules like CN in cool stars, we used the
spectrum-synthesis method, calculating five spectra having different
abundances  and adopting the one that minimizes the r.m.s.
The O content was obtained from the forbidden line at 6300 \AA ~ and Na
from the 6154\AA ~ line (and also the 6160\AA ~ line for warmer stars). The O line was decontaminated 
from telluric lines using  an O-type star. 
We were unable to measure O for one lower RGB star due to cosmic ray contamination.

Atmospheric parameters were obtained as follows. 
First, T$_{\rm eff}$ was derived from the B-V, V-I, V-J, V-H, V-K, J-H, and
J-K colors using the relations by \citet{Al99} and \citet{Ra05} and taking the mean.
Surface gravities (log(g)) were obtained from the canonical equation:

\begin{center}
{\rm log($\frac{g}{g_{\odot}}$) = log($\frac{M}{M_{\odot}}$) + 4$\cdot$
log($\frac{T_{\rm eff}}{T_{\odot}}$) - log($\frac{L}{L_{\odot}}$)}
\end{center}

The bolometric correction was derived from the relations 
of \citet{Al99} and \citet{Fl96}.
The reddening E(B-V), distance modulus ${(m-M)_{\rm V}}$, and mass were
obtained from isochrone fitting of the V vs. B-V 
CMD using BaSTI \citep{Pi04} isochrones (Fig.~\ref{f1}). We obtained E(B-V)=0.13, ${(m-M)_{\rm V}}$=13.44, 9 Gyr and 
a mass of 1.13 M$_{\odot}$ for RGB stars, and 1.05 M$_{\odot}$ for RC and AGB stars.
All of these are in good agreement with previous values,
e.g. \citet{Ca06}. In particular, the mass agrees well with the value of $1.087\pm 0.004M_{\odot}$
derived by \citet{Br11} for the turnoff mass from detailed observations of a binary. 
Microturbulent velocity (v$_{\rm t}$) was obtained from the
relation of \citet{Gr96} that utilizes both temperature and gravity:

\begin{center}
{\rm v$_{\rm t}$=0.00119$\cdot$T$_{\rm eff}$-0.90$\cdot$log(g)-2}
\end{center}

The input metallicity 
was  [Fe/H]=+0.40. 
The LTE program MOOG \citep{Sn73} was used
for the abundance analysis coupled with atmosphere models by \citet{Ku92}.
We adopted the same linelist used in our previous papers
(e.g. \citealt{Vi10}). Atmospheric parameters and final derived 
abundances are reported in Tab.~\ref{t2}, together with adopted solar values.

We performed a check on our atmospheric parameters by plotting 
  our targets and all cluster stars in log(g)  
and log(L/L$_{\odot}$) vs. T$_{\rm eff}$ diagrams in Fig.~\ref{f2}.
The RGB is well defined in both
diagrams, indicating that if any differential reddening is present \citep{Pl11}, it does not
affect our parameters significantly. 
The AGB and RGB are well separated,
allowing us to confidently assign the proper mass to each target. 
We compare our parameters with an appropriate BaSTI model of 9 Gyr
and find no difference in luminosity but a systematic difference in 
temperature of 120 K. Otherwise, all our stars are located on or
very close to the theoretical model, confirming the reliability of our procedure.

NLTE effects can influence Na abundance determinations. The lines we used are the least affected, and
the influence is minimal at this high metallicity \citep{Li11}. More importantly, differential effects are very
small over the range of parameters of our sample. \citet{Li11} show the maximum difference expected
is only 0.05 dex. Thus, we did not make any correction. 

\begin{figure*}
\centering
\includegraphics[angle=0,width=16cm]{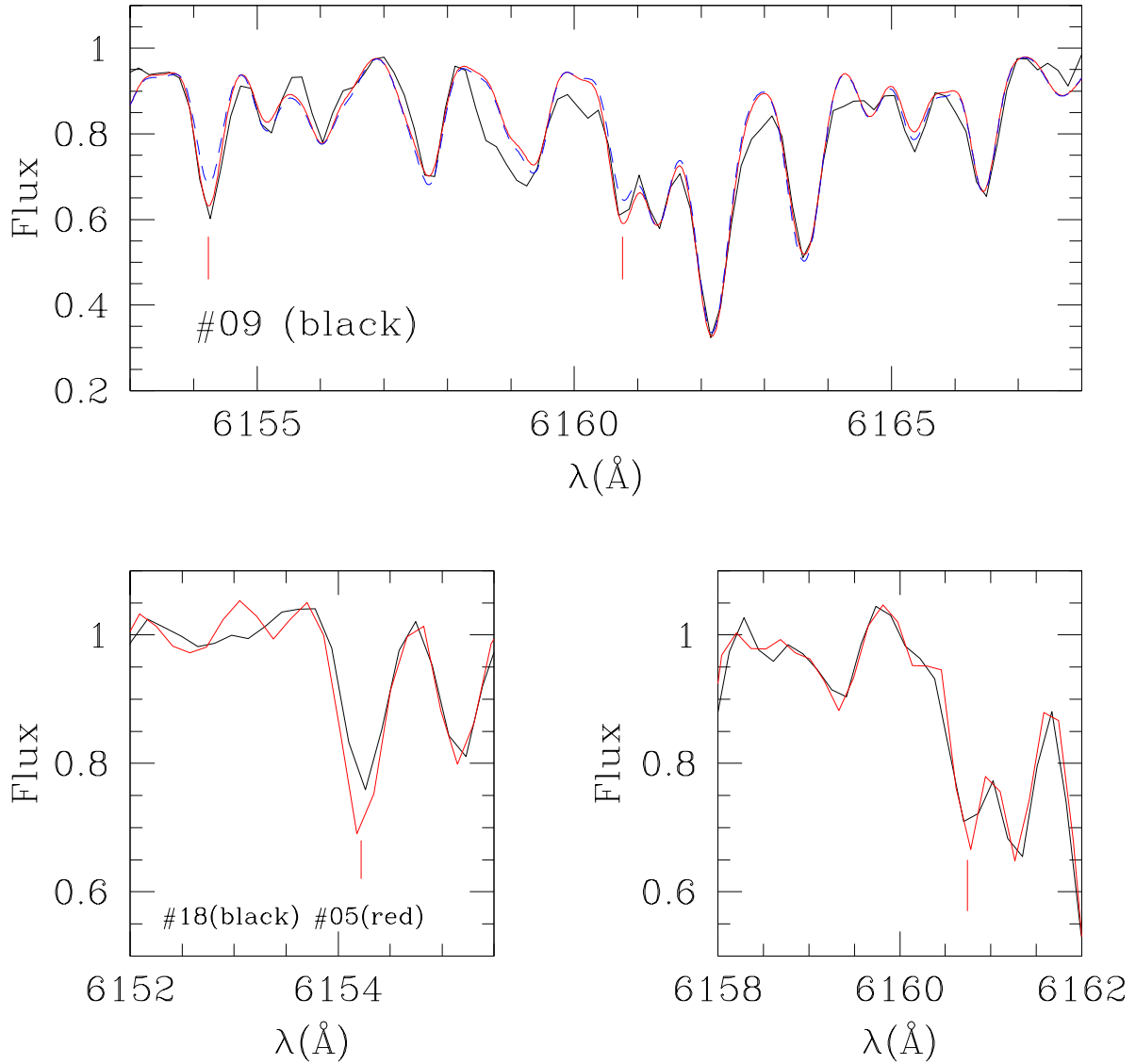}
\caption{Top: Spectrum  around the NaI 6154 and 6160\AA  ~ lines (marked) for the coolest star, together with a synthetic spectrum for [Na/Fe]=0 (blue) and +0.5 (red).
Bottom: Spectra of two RC stars in the same
region. The stars (T05 and T18) have almost
identical atmospheric parameters but a wide range in Na absorption strength is evident, confirming
the large difference in Na abundance. The 6160\AA ~ line is strongly blended and was not used 
in the abundance analysis but still shows the differential absorption.
}
\label{f3}
\end{figure*}

A sample of our stars exhibited an anticorrelation between our initial Na abundance and $T_{eff}$.
Following the referee's suggestion, we investigated if 
this could be due to blends with species such as CN not properly accounted for in the linelist.
We estimated the N abundance from                  the
   strength of the CN feature at 6195-6198\AA ~  in four stars covering the range
of Na abundance and $T_{eff}$, finding [N/Fe] =+0.2. We then obtained final 
Na abundances for all stars, using this N
abundance. The Na content was lowered by $\sim$0.1 dex in
all stars, with two of our coolest stars being most affected. 
Fig.~\ref{f3} includes a portion of the spectrum around the Na lines for our coolest star
together with the best-fit synthetic spectrum. Although some residual absorption may be present,
it minimally affects the abundance derived from the 6154\AA  ~ line, the only one used for these cool stars.

An internal error analysis was performed by varying T$_{\rm eff}$, log(g), [Fe/H], and
v$_{\rm t}$ by an amount equal to the estimated internal error and redetermining abundances of star \#T05, assumed to represent
the entire sample. Parameters were varied by $\Delta$T$_{\rm eff}$=+10 K,
$\Delta$log(g)=+0.05, $\Delta$[Fe/H]=+0.05 dex, and $\Delta$v$_{\rm t}$=+0.04
km/s. The temperature error was obtained by comparing the individual color-based
determinations for each star, while the errors in gravity and
microturbulence were obtained applying error propagation to the previous equations 
assuming an internal uncertainty of 0.05 M$_{\odot}$. 
The [Fe/H] error was taken as the r.m.s. of our results.
Other error sources such as uncertainties in the distance modulus and
reddening affect our results systematically and can be neglected here.
We stress the fact that these are only internal errors. Systematic errors are
certainly larger but not of major concern.
Total internal abundance errors ($\sigma_{\rm tot}$), including spectral noise, are 0.07,
0.05, and 0.05 dex for [O/Fe], [Na/Fe], and [Fe/H] respectively. 


\begin{figure*}
\includegraphics[angle=0,width=8cm]{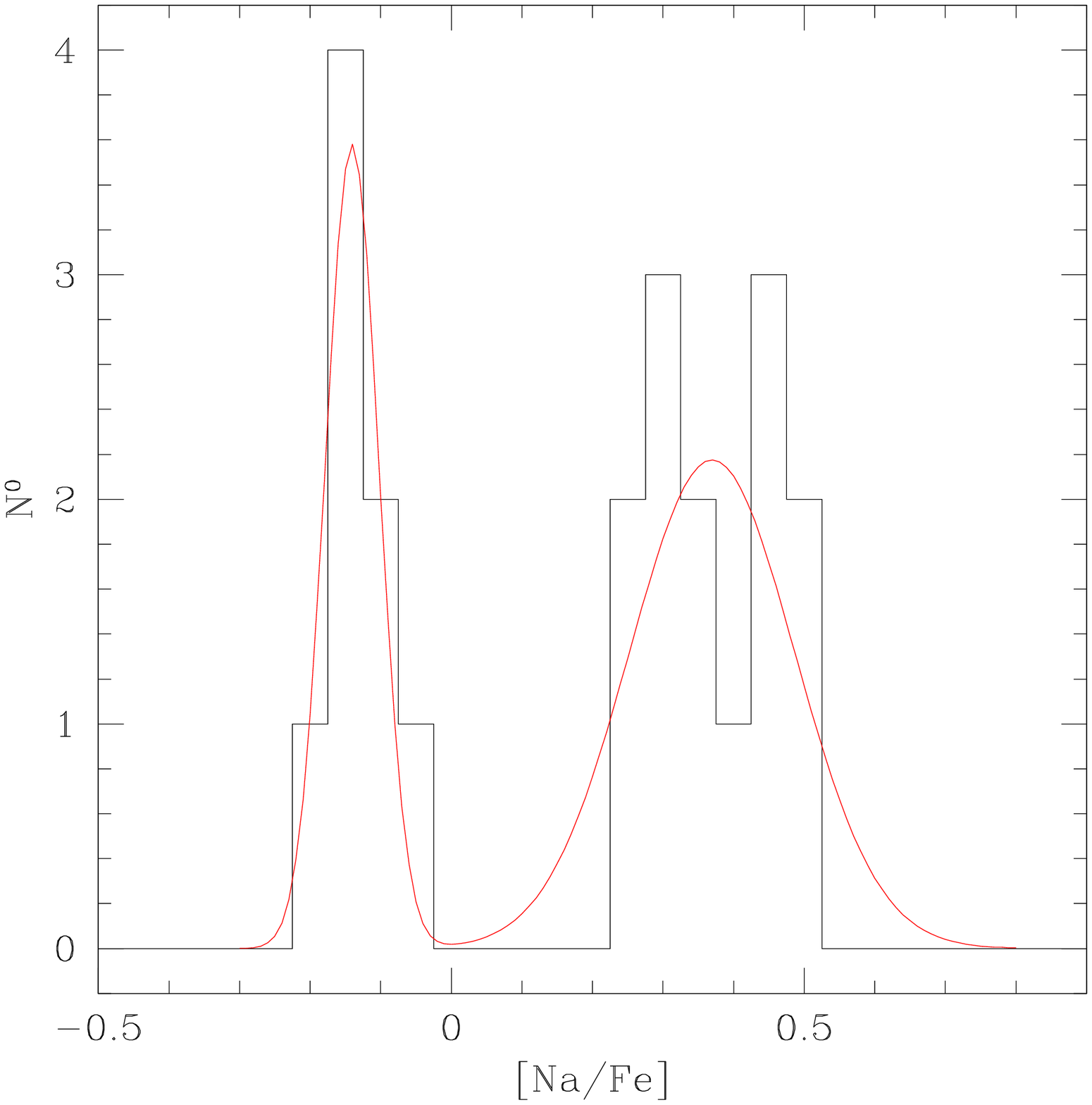}
\includegraphics[angle=0,width=8cm]{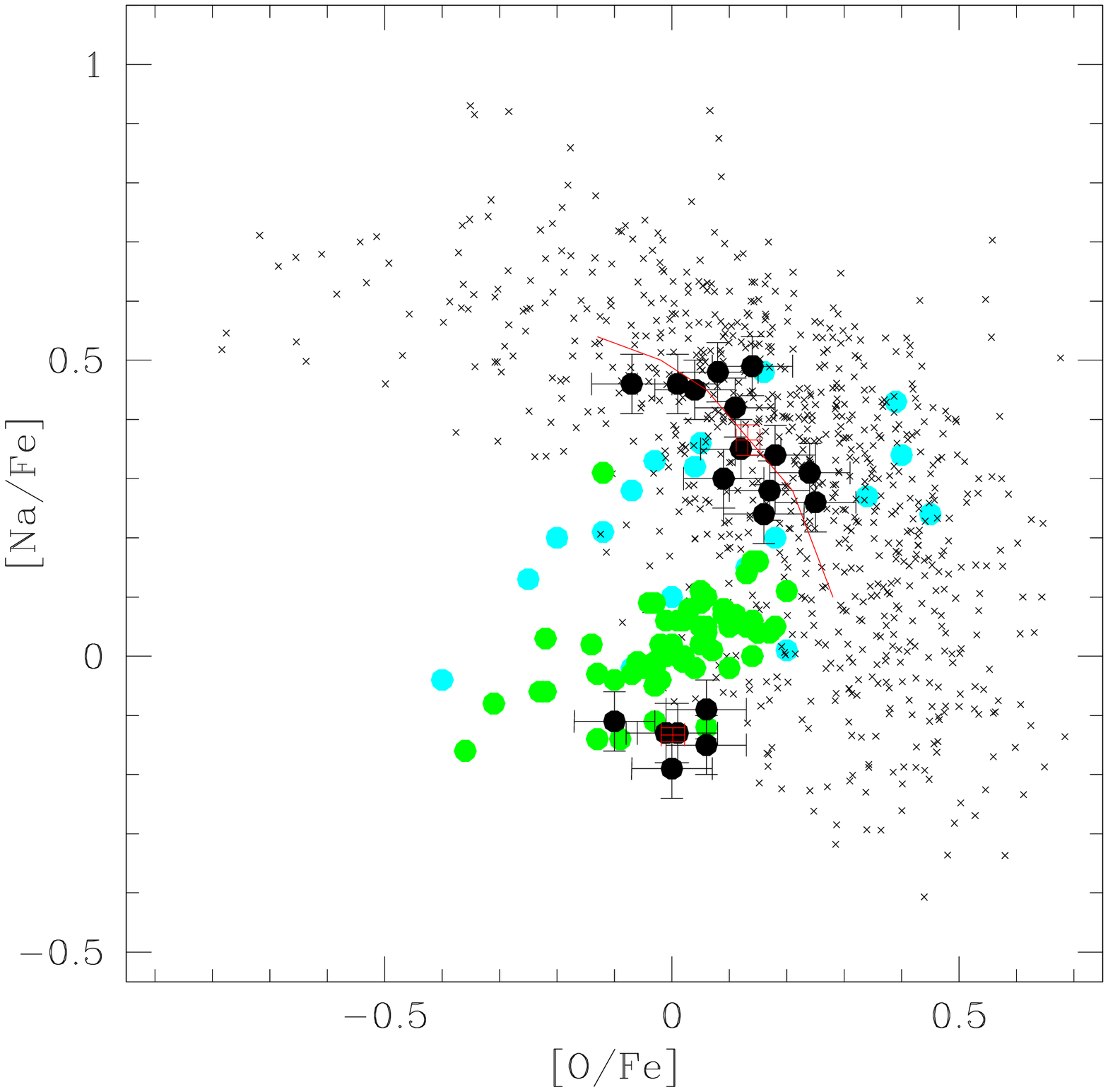}
\caption{Left: Histogram of the [Na/Fe] abundance ratio
distribution (lines) with a 2-Gaussian fit (curves). Right: 
[Na/Fe] vs. [O/Fe] for stars in NGC 6791 (filled circles with error
bars), GC stars (crosses), metal rich ([Fe/H]$>$-0.2) field stars (green filled circles)
and the means for OCs from \citet{De09} (blue filled circles). The mean GC anticorrelation is shown by the red curve.}
\label{f4}
\end{figure*}

\section{Results}

\subsection{Fe}


We found a mean [Fe/H]  of +0.42$\pm$0.01, 
in the middle of literature values.
\citet{Ca06} finds [Fe/H]=+0.39, \citet{Or06} +0.35,
while \citet{Ca07} gives +0.46. 
We have an independent confirmation of our value.
The HIRES spectra are analyzed in more detail in \citet{Vi12}, where the 
atmospheric parameters are  
derived purely spectroscopically.
Nevertheless,  \citet{Vi12} also obtain
[Fe/H]=+0.42$\pm$0.02. The perfect agreement between the observed dispersion, 0.05 dex, and that
from error analysis demonstrates that NGC 6791 lacks any measurable intrinsic
metallicity dispersion, in accord with all other studies and as expected for an OC.

\begin{table*}
\caption{Optical photometry, atmospheric parameters and abundances for the observed stars.}            
\label{t2}      
\centering
\begin{tabular}{lcccccccll}        
\hline\hline
ID & B(mag) & V(mag) & I(mag) &
 T$_{\rm eff}$(K) & log(g) & v$_{\rm t}$(km/s) & [Fe/H] & [O/Fe] & [Na/Fe]\\
\hline             
T01 & 15.817 & 14.482 & 13.146 & 4447 & 2.30 & 1.22 & 0.40 &  0.06 & -0.09\\
T03 & 15.923 & 14.588 & 13.264 & 4429 & 2.33 & 1.17 & 0.49 & -0.10 & -0.11\\
T04 & 16.084 & 14.665 & 13.182 & 4195 & 2.21 & 1.00 & 0.37 &  0.24 &  0.31\\
T05 & 15.874 & 14.546 & 13.235 & 4465 & 2.34 & 1.21 & 0.41 &  0.18 &  0.34\\
T06 & 15.841 & 14.529 & 13.168 & 4400 & 2.29 & 1.18 & 0.41 &  0.12 &  0.35\\
T07 & 15.357 & 13.741 & 11.962 & 3951 & 1.57 & 1.29 & 0.39 &  0.09 &  0.30\\
T09 & 16.160 & 14.713 & 13.230 & 4226 & 2.26 & 1.00 & 0.47 &  0.14 &  0.49\\
T10 & 15.424 & 13.849 & 12.191 & 4033 & 1.70 & 1.27 & 0.37 &  0.11 &  0.42\\
T11 & 15.949 & 14.459 & 12.890 & 4163 & 2.10 & 1.06 & 0.39 &  0.17 &  0.28\\
T12 & 16.032 & 14.557 & 13.027 & 4184 & 2.16 & 1.04 & 0.47 &  0.04 &  0.45\\
T13 & 16.016 & 14.602 & 13.274 & 4439 & 2.35 & 1.17 & 0.49 &  0.16 &  0.24\\
T14 & 15.904 & 14.551 & 13.217 & 4427 & 2.32 & 1.18 & 0.38 &  0.25 &  0.26\\
T15 & 15.729 & 14.136 & 12.373 & 3942 & 1.75 & 1.12 & 0.36 &  0.08 &  0.48\\
T17 & 16.059 & 14.554 & 12.988 & 4119 & 2.10 & 1.01 & 0.46 &  0.01 &  0.46\\
T18 & 15.874 & 14.515 & 13.176 & 4468 & 2.33 & 1.22 & 0.40 &  0.01 &  -0.13\\
T19 & 15.495 & 13.862 & 11.892 & 3822 & 1.48 & 1.22 & 0.53 & -0.07 &  0.46\\
T31 & 18.329 & 17.150 & 15.954 & 4699 & 3.56 & 0.39 & 0.40 &  0.00 & -0.19\\
T32 & 18.368 & 17.158 & 15.923 & 4672 & 3.55 & 0.37 & 0.45 &  0.06 & -0.15\\
T33 & 18.575 & 17.457 & 16.330 & 4894 & 3.79 & 0.42 & 0.38 & -0.01 & -0.13\\
T34 & 18.553 & 17.372 & 16.164 & 4727 & 3.67 & 0.33 & 0.40 &   -   &  -0.07\\
T35 & 18.520 & 17.370 & 16.210 & 4800 & 3.70 & 0.38 & 0.41 & -0.01 & -0.13\\
\hline                                   
\hline
Sun & - & - & - & 5777 & 4.44 & 0.80 &  7.50 &  8.80\tablenotemark{a}  &  
6.32\tablenotemark{b}\\
\hline
\end{tabular}
\tablenotetext{a}{~log(O/H)+12}
\tablenotetext{b}{~log(Na/H)+12}
\end{table*}

\subsection{O and Na}

However, our  O and Na analysis yields several completely surprising
results. First, [Na/Fe] has an observed dispersion of 0.26 dex,  $>5$
times larger than that expected from error analysis, while
that of [O/Fe], 0.10 dex,  is 1.5 times larger. 
Fig.~\ref{f3} compares the spectra of two different stars at the 
Na lines. Both are RC stars and have virtually identical atmospheric parameters, but exhibit a large 
variation in Na absorption and thus must have a large Na abundance difference.  
To further investigate this point,  Fig.~\ref{f4} (left) displays a
histogram of the Na distribution.
Two well separated populations appear. A KMM mixture-modelling test
\citep{As94} strongly supports a bimodal Gaussian
over a single-Gaussian distribution, at a confidence level of
$>99\%$.
The two-Gaussian fit to the distribution finds
one population with a mean and  dispersion of
[Na/Fe]=-0.14$\pm$0.02, $\sigma_{\rm [Na/Fe]}$=0.04$\pm$0.01, while the second
has [Na/Fe]=+0.36$\pm$0.03, $\sigma_{\rm [Na/Fe]}$=0.12$\pm$0.02. Thus, while
the dispersion of the first peak
is small, within the observational errors, the
dispersion of the second is $>2$ times larger than expected and appears quite significant. The
implication is that the second subpopulation has an intrinsic Na dispersion,
while the first is homogeneous. The Na-poor 
population includes all of the lower RGB stars as well as three RC stars, while
the other population is composed of RC stars as well as upper RGB and AGB stars. 
Thus, while evolutionary effects may be
involved, they cannot fully explain the observed behavior. 
In any case, Na and O are not predicted to be affected by evolutionary mixing in a significant way at this mass and 
metallicity \citep{Gr00}.


We are aware of a similar, unpublished study of Na and O abundances in NGC 6791. \footnote {A preliminary
report is available at $http://www.sexten-cfa.eu/public/2011/ChemEvoMilkyway/bragaglia.pdf$.} Briefly,
they use WIYN + Hydra to investigate a similar number of RC and RGB stars and find no evidence
for any abundance spread. However, their mean S/N is only 23 and thus their errors are much larger
than ours. While they interpret their result as a homogeneous composition
with a spread due only to errors, our more precise results allow us
to disentangle the two sub-populations and reveal the intrinsic variation. 

The real nature of the two populations is revealed when we plot [Na/Fe]
vs. [O/Fe] (Fig.~\ref{f4}, right panel). Our data are compared with
the database on GCs by \citet{Ca09} and metal rich ([Fe/H]$>$-0.2) field stars
from \citet{Re03,Re06}. The Na-poor population is well
separated from the GC trend, with a mean O content and dispersion of 
[O/Fe]=+0.00$\pm$0.02,  $\sigma_{\rm [O/Fe]}$=0.05$\pm$0.01, 
while the Na-rich population has
[O/Fe]=+0.13$\pm$0.02,  $\sigma_{\rm [O/Fe]}$=0.07$\pm$0.01. 
The mean O and Na contents with their errors are shown.
The significance of the difference between the two [Na/Fe] subpopulations is 16
$\sigma$,  and 5 $\sigma$ for [O/Fe]. Thus, it appears
that there is a real spread (perhaps bimodality) in O as well as Na.
The Na-poor population shows a homogeneous Na and O content, similar to field stars, and no trend
appears. 
The distribution of Na-O abundances for the Na-rich population nicely follows part of the mean
GC Na:O anticorrelation. 

The above behavior is extraordinary in several ways. First, NGC 6791 becomes the
first OC to display an intrinsic dispersion in any element that is unlikely to be
explained by mixing effects, and therefore the first (presumed) OC discovered with multiple populations. Na shows a clear spread, while the spread in O is not
as strong but still likely. Secondly, Na exhibits bimodality. Thirdly, the
Na-rich population also appears to have an internal spread.
Finally, this
population follows the almost  ubiquitous Na:O anticorrelation seen in  GC giants. As
such, the \citet{Ca09} GC definition implies that at least this 
population of NGC 6791 stars constitute a GC!? If one believes the minimum mass limits so far 
derived in order to form multiple populations, then NGC 6791 must have lost at least 90\% of its 
original mass. This is in agreement with expectations for ``normal" GCs (D'Ercole et al. 2008).   We also find more ``second generation" or Na-enhanced stars than Na-poor stars, which is also expected from GC formation models and seen in other GCs \citep{Ca10}.


The above raises the fundamental questions: what is NGC 6791 and what was its origin? Is it an OC, 
as always considered; a GC, as suggested by its Na:O anticorrelation; a hybrid, or some other type of
unique object? It is so far the only supposed OC to show multiple populations.
Note that the Na-poor population overlaps reasonably well with disk field stars but not with other OCs, while the Na-rich
population falls along the mean OC trend. 
Clearly, the formation of such a peculiar object was complex and requires new ideas. NGC~6791's present-day mass ($\sim 5\times$10$^3$ M$_{\odot}$, \citet{Ki65,Or06,Pl11}) 
is far below the predicted minimum mass needed to retain gas and form a second
generation, but its initial mass could have been much larger. 

Perhaps, as proposed by \citet{Ca06,Bu12}, NGC~6791 is the remnant of
a dwarf galaxy captured and tidally disrupted by the Milky Way. In this case the two
sub-populations might have formed as independent clusters, one presumably much more massive and GC-like, and then
merged in the core of the host galaxy and survived disruption. 
\citet{Tw11} find evidence for a radial age spread of $\sim$1 Gyr,
further substantiating the suggestion of multiple star formation epochs.
However, we see no significant difference in the radial distribution, mean velocity or its dispersion
of the two Na subpopulations.
As \citet{Ca06}
point out, we are also left with the major problem of explaining the formation of extremely metal-rich stars,
which would normally require a very massive environment, many orders of magnitude larger than the
current mass.
What is clear is that NGC 6791 is neither a traditional OC nor GC but an extraordinary 
object with much left to explore and reveal to us. Similar observations of other massive OCs and low
mass GCs would be of great interest.

\section{Conclusion}

We analyzed high resolution, high S/N spectra from two independent datasets for 21 member stars covering a wide range of evolutionary state in the traditional OC
NGC~6791. We obtained O, Na, and Fe abundances with small internal errors.
We found a homogeneous [Fe/H]=+0.42$\pm 0.01$.
Surprisingly, stars are divided into two subpopulations with different mean O and especially
Na contents. The significance of these differences are many $\sigma$. Thus, NGC 6791 
becomes the first OC to display an intrinsic dispersion in any element and the first presumed OC discovered 
with multiple populations.
It is also the first cluster of any kind to show 
Na-poor stars with a homogeneous Na content, along with a Na-rich group showing an
intrinsic Na spread. The Na-poor group falls near the field star O/Na content,
while the Na-rich population follows the Na-O anticorrelation typical of
GCs. NGC 6791 defies the traditional definition of either an OC or GC.
How such a complex and highly enriched object was formed is unknown.

\acknowledgments
D.G. and S.V. gratefully acknowledge support from the Chilean 
project BASAL   Centro de Excelencia en Astrof\'isica
y Tecnolog\'ias Afines (CATA) grant PFB-06/2007. 
This material is based upon work supported by the National Science Foundation under award No. ASTÐ1003201 to CIJ. We thank an anonymous referee for significant contributions.

\end{document}